\documentclass[english,aps,prb,twocolumn,10pt]{revtex4-1}
\usepackage{lmodern}
\usepackage[T1]{fontenc}
\usepackage[latin9]{inputenc}
\usepackage{babel}
\usepackage{units}
\usepackage{amsmath}
\usepackage{amssymb}
\usepackage{graphicx}
\usepackage{bm}
\usepackage{xcolor}
\usepackage{soul}

\newcommand{\comment}[1]{}

\makeatother

\begin{document}
\title{Microscopic origin of the bolometric effect in graphene}

\author{Roland Jago}
\author{Ermin Malic}
\author{Florian Wendler}
\affiliation{Chalmers University of Technology, Department of Physics, SE-412 96 Gothenburg, Sweden}
\begin{abstract}
 While the thermoelectric and photoconduction effects are crucial in pristine and low-doped graphene, the bolometric effect is known to dominate the photoresponse in biased graphene. Here, we present a detailed microscopic investigation of the photoresponse due to the bolometric effect in graphene. Based on the semiconductor Bloch equations, we investigate the time- and momentum-resolved carrier dynamics in graphene in the presence of a constant electric field under optical excitation. The magnitude of the bolometric effect is determined by the optically induced increase of temperature times the conductivity change. Investigating both factors independently, we reveal that the importance of the bolometric effect in the high-doping regime can be mostly ascribed to the latter showing a parabolic dependence on the doping.
\end{abstract}
\maketitle

%====================================================================================================================================================
\subsection*{Introduction}
\begin{figure}[!t]
\begin{centering}
\includegraphics[width=1\columnwidth]{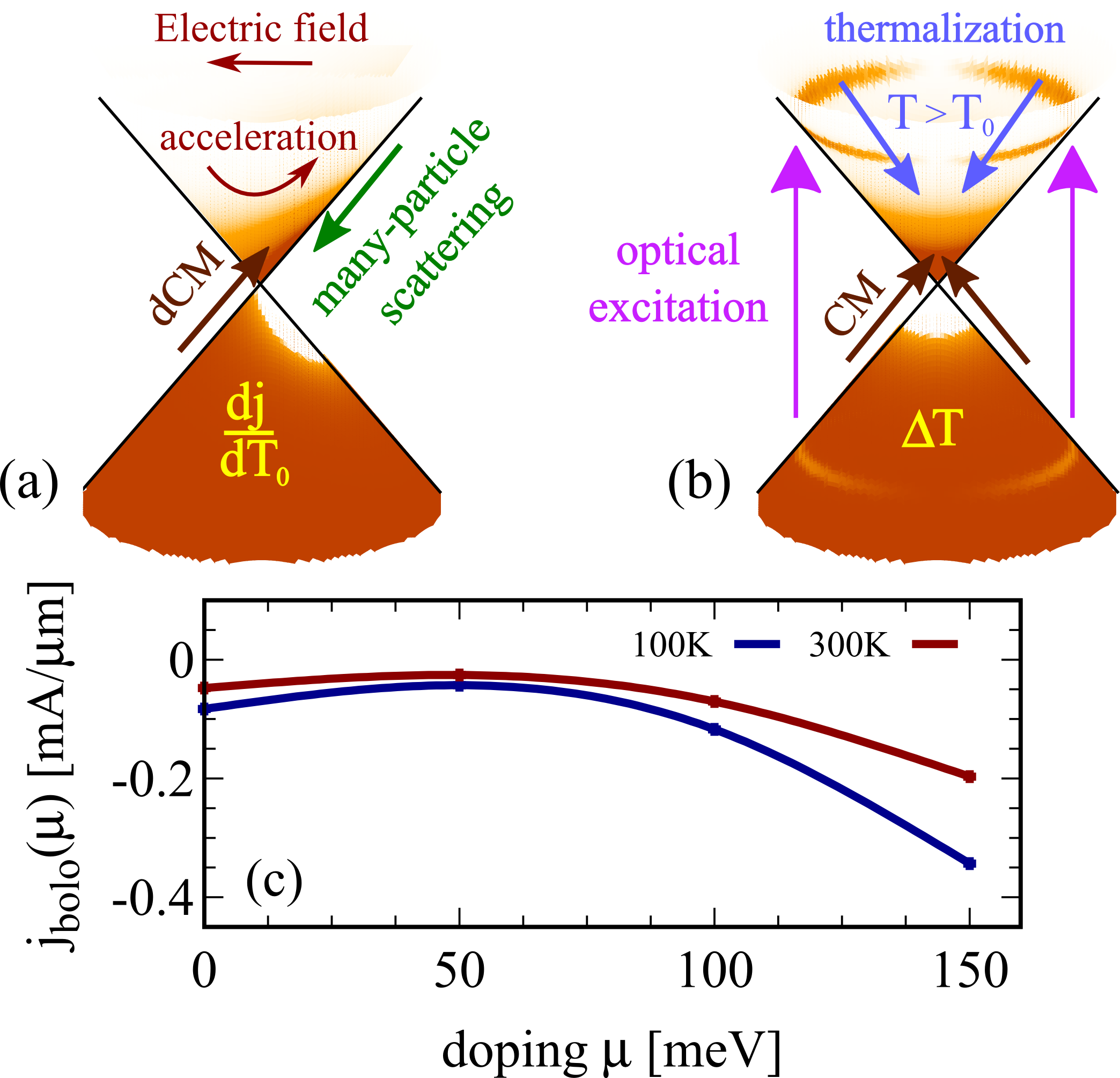} 
\par\end{centering}
\caption{(a) An electric field shifts the carriers in reciprocal space, whereas many-particle scattering counteracts this field-induced acceleration and brings the hot carriers back to the Dirac point. Due to the linear band structure of graphene, Auger scattering is efficient and results in a dark carrier multiplication (dCM) increasing the charge carrier density \cite{Jago_darkCM_2017}. (b) Optical excitation without applied electric field: Photo-excited carriers thermalize to the Dirac point and form a hot Fermi distribution. (c) Bolometric photocurrent for two different temperatures in dependence of doping.}
\label{fig:sketch} 
\end{figure}

%====================================================================================================================================================
Graphene-based photodetectors have attracted a great deal of interest in recent years \cite{Koppens2014,sun2014,Buscema2015}, since graphene absorbs light over a broad spectral range and its ultrafast carrier dynamics facilitates the operation of photodetectors at unprecedented speeds \cite{Mueller2010, echtermeyer11, furchi12,ferrari10, bonaccorso10, avouris12, Malic2011}. This could be the key for satisfying the growing need for fast optical data transfer in modern technology. Due to the superposition of different processes that are at play when light is shined on graphene, the origin of the photoresponse is not always clear and it is difficult to make predictions. Important mechanisms for photodetection in graphene are (i) the bolometric effect where optical excitation induces a heating which changes the conductivity and the corresponding current, (ii) the thermoelectric effect which uses the Seebeck effect to transform the optically induced heat into a current, and (iii) the photoconduction effect in which an external electrical field accelerates opticaaly excited carriers leading to a current. While the photoresponse in pristine graphene is mainly attributed to the thermoelectric and the photoconduction effects\cite{Xu2010,Gabor2011,Freitag2013,Echtermeyer2014}, it has been shown that the bolometric effect is the dominating mechanism in biased graphene \cite{Freitag2013a,Mittendorff2013_thzdetector,Mittendorff2015_Photodetector,Schuler2016}. Exploiting these findings, Mittendorff et al. have demonstrated an ultrafast graphene-based broadband THz detector working at room temperature \cite{Mittendorff2013_thzdetector}.
Here, we shed light on the microscopic mechanism behind photodetection based on the bolometric effect in graphene. Even in a microscopic theory, it is not trivial to discern the photoconduction effect from the bolometric effect, since both effects occur at the same time and rely on an enhanced carrier density. The linear band structure of graphene gives rise to nonlinear optical response and additional contributions to the conductivity, see Ref. \cite{Ganichev2014}. For example the optical excitation generates an anisotropic distribution of optically excited carriers. To determine the bolometric effect we separate the influence of the optical excitation from the transport current.
The complex carrier dynamics, i.e. the time- and momentum-resolved carrier distribution, in optically excited graphene subjected to an in-plane electric field has been investigated in a previous work \cite{Jago_photo2017}. To reveal the microscopic foundations of the bolometric effect, our strategy is to independently investigate the two contributions responsible for the bolometric photocurrent: (i) The change of the conductivity with temperature $ dj/dT_{0}$ (Fig. \ref{fig:sketch}(a)), and (ii) the temperature change $\Delta T$ induced by an optical excitation  (Fig. \ref{fig:sketch}(b)).
Figure \ref{fig:sketch}(c) shows the predicted bolometric photocurrent density  $ dj/dT_{0}\cdot \Delta T$ in dependence of doping for two different temperatures. Overall, the bolometric photocurrent is the largest at high doping values and at small temperatures. 

%====================================================================================================================================================
\subsection*{Theoretical approach}
%====================================================================================================================================================
To obtain microscopic access to the dynamics of carriers in the presence of an electrical field, we derive a set of coupled equations of motion for the microscopic quantities: the electron/hole occupation probability $\rho_{\mathbf{k}}^{\tilde{\lambda}}=\langle a_{\mathbf{k}\tilde{\lambda}}^{\dagger}a_{\mathbf{k}\tilde{\lambda}}^{\phantom{\dagger}}\rangle$,
the microscopic polarization $p_{\mathbf{k}}=\langle a_{\mathbf{k}v}^{\dagger}a_{\mathbf{k}c}^{\phantom{\dagger}}\rangle$, and the phonon number $n_{\mathbf{q}}^{j}=\langle b_{\mathbf{q}j}^{\dagger}b_{\mathbf{q}j}^{\phantom{\dagger}}\rangle$. Here, the creation and annihilation operators $a_{\mathbf{k}\tilde{\lambda}}^{\dagger}$ and $a_{\mathbf{k}\tilde{\lambda}}^{\phantom{\dagger}}$ with momentum $\mathbf{k}$ are used for electrons in the valence or conduction band ($\tilde{\lambda} = v, c$), while the hole occupation probability is given by $\rho_{\mathbf{k}}^{h} = 1 - \rho_{\mathbf{k}}^{v}$. The corresponding phonon operators are $b_{\mathbf{q}j}^{\dagger}$, $b_{\mathbf{q}j}^{\phantom{\dagger}}$, with phonon mode $j$ and phonon momentum $\mathbf{q}$. 

The derivation of the equations requires knowledge of the  many-particle Hamilton operator $H$. In this work, we take into account the (i) free carrier and phonon contribution $H_{\text{0}}$, (ii) carrier-carrier $H_{\text{c-c}}$ and (iii) carrier-phonon $H_{\text{c-ph}}$ interaction accounting for Coulomb- and phonon-induced scattering, (iv) the carrier light coupling $H_{\text{c-l}}$ that is treated on a semi-classical level, and (v) finally $H_{\text{c-f}}$ describing the interaction of an external electric field $\bf{E}$ with electrons. The Hamilton operator for the electron-field interaction is given by 
\begin{align}
H_{\text{c-f}} & =-ie_{0}\mathbf{E}\cdot\sum_{\mathbf{k}\tilde{\lambda}}a_{\mathbf{k}\tilde{\lambda}}^{\dagger}\nabla^{\phantom{\dagger}}_{\mathbf{k}}a^{\phantom{\dagger}}_{\mathbf{k}\tilde{\lambda}},\label{eq:H_cf}
\end{align}
with the elementary charge $e_{0}$. Details on the other contributions of the many-particle Hamilton operator can be found in Refs. \onlinecite{Malic, Malic2011, Binder}. Exploiting the Heisenberg equation, we derive the graphene Bloch equations in second-order Born-Markov approximation yielding
\begin{eqnarray}
\dot{\rho}_{\mathbf{k}}^{\lambda} & =&\Gamma_{\mathbf{k}\lambda}^{\text{in}}\,\big(1-\rho_{\mathbf{k}}^{\lambda}\big)-\Gamma_{\mathbf{k}\lambda}^{\text{out}}\,\rho_{\mathbf{k}}^{\lambda}\nonumber\\
&&+2\,\text{Im}\left(\Omega_{\bf k}^{vc, *} p_{\bf{k}}\right)
-\frac{e_{0}}{\hbar}\mathbf{E}\cdot\nabla_{\mathbf{k}}\rho_{\mathbf{k}}^{\lambda},\label{eq:rho}\\[5pt]
\dot{p}_{\bf{k}} & =&i\Delta\omega_{\bf{k}}p_{\bf{k}}-i\Omega_{\bf k}^{vc}\left(\rho_{\bf{k}}^{e}+\rho_{\bf{k}}^{h}-1\right)-\frac{e_{0}}{\hbar}\mathbf{E}\cdot\nabla_{\mathbf{k}}p_{\mathbf{k}},\label{eq:p}\\[5pt]
\dot{n}_{\mathbf{q}}^{j}& =&\Gamma_{\mathbf{q}j}^{\text{em}}\,\big(n_{\mathbf{q}}^{j}+1\big)-\Gamma_{\mathbf{q}j}^{\text{abs}}\, n_{\mathbf{q}}^{j}-\gamma_{\text{ph}}\,\big(n_{\mathbf{q}}^{j}-n_{\mathbf{q},\text{B}}^{j}\big).\label{eq:n}
\end{eqnarray}
The equations describe the time- and momentum-resolved coupled dynamics of electrons/holes ($\lambda=e,h$), phonons, and the microscopic polarization. The dynamics of electrons in the conduction band and holes in the valence band are symmetric, but have different initial conditions for doped graphene samples.
Note that the sign of the field term is the same for electrons and holes, meaning that electrons and holes move in the same direction in $\mathbf{k}$-space, but in different directions in real space, since the group velocities are given by $\mathbf{v}^{e}_{\mathbf{k}} = +v_{F}\mathbf{k}/|\mathbf{k}|$ and $\mathbf{v}^{h}_{\mathbf{k}} = -v_{F}\mathbf{k}/|\mathbf{k}|$ with Fermi velocity $v_{F}$, for a detailed discussion see Ref. \onlinecite{Jago_darkCM_2017}.
The appearing Rabi frequency is defined as $\Omega_{\bf k}^{vc}(t)=i\frac{e_{0}}{m_{0}}  \mathbf{M}_{\bf k}^{vc}\cdot \mathbf{A}(t)$ with the free electron mass $m_{0}$, the vector potential $\mathbf{A}(t)$, and the optical matrix element $\mathbf{M}_{\mathbf{k}}^{vc}=\langle\mathbf{k}v|\nabla_{\mathbf{k}}|\mathbf{k}c\rangle$. Since we study the carrier dynamics close to the Dirac point renormalization effects can be neglected.
Furthermore, we have introduced $\hbar\Delta\omega_{\mathbf{k}}(t)=(\varepsilon_{\mathbf{k}}^{v}-\varepsilon_{\mathbf{k}}^{c} + i\gamma_{\mathbf{k}}(t))$ with the electronic dispersion $\varepsilon_{\mathbf{k}}^{\tilde{\lambda}}$ and the dephasing rate $\gamma_{\bf k}(t)$. The time- and momentum-dependent dephasing  $\gamma_{\mathbf{k}}(t)$ and in- and out-scattering rates $\Gamma_{\mathbf{k}\lambda}^{in/out}(t)$ include carrier-carrier and carrier-phonon scattering channels. The dynamics of the phonon number $n_{\mathbf{q}}^{j}(t)$ is driven by the emission and absorption rates  \cite{Malic,Malic2011} $\Gamma_{\mathbf{q}j}^{em/abs}(t)$. Here, $n_{\mathbf{q},\text{B}}^{j}$ is the initial Bose-distribution for phonons and $\gamma_{\text{ph}}$ is the experimentally determined phonon decay rate \cite{Kang2010}. For more details on the appearing many-particle scattering and dephasing rates, see Refs.  \onlinecite{Malic, Malic2011}. 

Transforming the system to a moving reference frame via the coordinate transformation $\mathbf{k}\rightarrow\mathbf{k}-\frac{e_{0}}{\hbar}\mathbf{E}t$ and $\frac{d}{dt}\rightarrow\frac{d}{dt}-\frac{e_{0}}{\hbar}\mathbf{E}\cdot\nabla_{\mathbf{k}}$, the field terms in the Bloch equations (\ref{eq:rho})-(\ref{eq:n}) disappear \cite{Meier1994}. Then, these equations correspond to the standard graphene Bloch equations without electric field, where the dynamics induced by the field is hidden in the motion of the coordinates. Therefore, applying an electric field induces a shift of $\rho_{\mathbf{k}}^{\lambda}$ and $p_{\mathbf{k}}$ in the reciprocal space. The reason why we need to perform this transformation although we are considering quasi-stationary solutions in the end is that these solutions emerge from a complex carrier dynamics which needs to be considered in order to arrive at the correct solution. Consider, for instance, the case of an optical excitation followed by a thermalization of optically excited charge carriers towards a quasi-equilibrium carrier distribution happening on a 100fs timescale \cite{Malic}. While the quasi-equilibrium carrier distribution depends on the carrier multiplication during the process of thermalization, the carrier multiplication vanishes as soon as the quasi-equilibrium is reached.

Numerically solving the Bloch equations (Eqs. (\ref{eq:rho})-(\ref{eq:n})) gives a full microscopic access to the time- and energy resolved carrier and phonon dynamics under the influence of an electric field. The field-induced shift of the carrier occupation in the linear band structure induces an acceleration of carriers. This leads to an asymmetric carrier distribution and results in a current \cite{Jago_photo2017}
\begin{align}
{\bf{j}}(t) & = -\frac{g\, e_{0}v_{\text{F}}}{A}\sum_{\bf{k}\lambda}\rho^{\lambda}_{\bf k}(t)\,\bf{e}_{\bf{k}},\label{eq:current}
\end{align}
with the sample area $A$, the unity vector $\mathbf{e}_{\mathbf{k}} = \mathbf{k}/|\mathbf{k}|$ and the degeneracy factor $g$, which equals $4$ by taking spin and valley degeneracy into account. The sum contains both types of carriers, namely electrons in the conduction band and holes in the valence band.

In this work, we assume that graphene lies on a $\text{SiC}$-substrate  and is surrounded by air on the other side. This is taken into account by introducing an averaged dielectric background constant\cite{Patrick1970}
$\varepsilon_{bg}=\frac{1}{2}\big(\varepsilon_{\text{s}}+1\big)$, where $\varepsilon_{\text{s}}= 9.66$ is the static screening constant of the substrate, while 1 describes the dielectric constant of air. Furthermore, the internal many-particle screening is determined by the static Lindhard formula \cite{Kira2006,Koch}
\begin{align}
  \varepsilon(\mathbf{q},t) = 1 - 2\,\frac{V_{\mathbf{q}}}{\varepsilon_{bg}}\sum_{\mathbf{k},\lambda\lambda'}\frac{\rho^{\lambda}_{\mathbf{k}}(t) - \rho^{\lambda'}_{\mathbf{k-q}}(t)}{\varepsilon^{\lambda}_{\mathbf{k}} - \varepsilon^{\lambda'}_{\mathbf{k-q}}}\big|C^{\lambda\lambda'}_{\mathbf{q}}(\mathbf{k})\big|^{2},
\end{align}
with the Coulomb potential $V_{\mathbf{q}}$ and the  prefactor $C^{\lambda\lambda'}_{\mathbf{q}}(\mathbf{k}) = \int d\mathbf{r}\Psi^{*}_{\lambda}(\mathbf{k})e^{-i\mathbf{q}\mathbf{r}}\Psi^{\phantom{*}}_{\lambda'}(\mathbf{k})$ with the tight-binding wave functions $\Psi_{\lambda}(\mathbf{k})$. The time dependence of the carrier distribution $ \rho^{\lambda}_{\mathbf{k}}(t)$ is explicitly taken into account by solving the graphene Bloch equations. In doped graphene the screening, especially for small transfer momenta $\mathbf{q}$, is enhanced.

The bolometric photocurrent\cite{Buscema2015} $j_{bolo}$ is given by the change of the quasi-equilibrium conductivity $\sigma$ with respect to the initial temperature $d\sigma/dT_{0} = d^{2}j/(dT_{0}dE)$ multiplied by the optically-induced temperature change $dT$ and applied electric field $E$. Since the current scales linearly with the electric field it is sufficient to consider $dj/dT_{0}$ instead of $Ed\sigma/dT_{0}$. Then, one obtains the bolometric photocurrent via integration
\begin{align}
 j_{bolo} = \int^{T_{0}+\Delta T}_{T_{0}}dT \frac{dj(T)}{dT_{0}} \approx \frac{dj(T_{0})}{dT_{0}}\cdot\Delta T. \label{eq:curr_bolo}
\end{align}
The approximated result is justified as long as $dj(T)/T_{0}$ is nearly constant in the interval $[T_{0},T_{0}+\Delta T]$.
To distinguish the bolometric effect from the photoconduction effect, we separately consider the two factors contributing to $j_{bolo}$ in Eq. \ref{eq:curr_bolo}: (i) the derivative $dj(T_{0})/dT_{0}$ calculated for the case without an optical excitation and (ii) the optically-induced temperature change $\Delta T$ calculated for the case without an electric field.

The first contribution, i.e. the temperature dependence of the conductivity, is obtained by determining the current density of graphene in a constant electric field at different temperatures. To this end, we exploit a theory based on the graphene Bloch equations \cite{Jago_darkCM_2017} capturing the microscopic carrier dynamics due to the carrier-field, carrier-carrier and carrier-phonon interactions acting on a femto- to picosecond timescale. The electric field induces a shift of thermally excited charge carriers away from the Dirac point, while many-particle scattering brings them back and thereby introduces a resistivity \cite{Jago_darkCM_2017}. Moreover, efficient Auger scattering in graphene \cite{Winzer2010_Multiplication,Brida2013,02_Wendler_CM_NatCommun_2014, Winzer2012_prb_rapid, Kadi2015,Ploetzing2014,05_Mittendorff_Auger_NatPhys_2014, Gierz2015} gives rise to an amplification of the carrier density that we denote as dark carrier multiplication (dCM), i.e. CM in the absence of an optical excitation. Here, we focus on the microscopic description of biased graphene whose photoresponse is governed by the bolometric effect.

The second contribution to the bolometric photoresponse $\Delta T$ is determined using the well-established carrier dynamics calculations of optically excited graphene \cite{AnnalenDerPhysik2017,Malic,Malic2011}. The optical excitation creates electron-hole pairs with energies matching the energy of the absorbed photons. The many-particle scattering brings the hot charge carriers back to the vicinity of the Dirac points, where they form a hot thermalized carrier distribution. This happens on a timescale of tens of femtoseconds, and is supported by Auger scattering, which is accompanied by carrier multiplication. Then, on a slower timescale of several picoseconds, the photo-excited charge carriers recombine while transferring their energy to the underlying lattice in form of phonons. With regard to ultrafast bolometric photodetectors, we focus on the photo-induced electron temperature increase, i.e. we determine the temperature after optical excitation by fitting a Fermi-Dirac distribution to the thermalized hot carrier distribution.
%====================================================================================================================================================
\subsection*{Temperature dependent dark current}
%====================================================================================================================================================
\begin{figure}[!t]
\centering
\includegraphics[width=1.05\columnwidth]{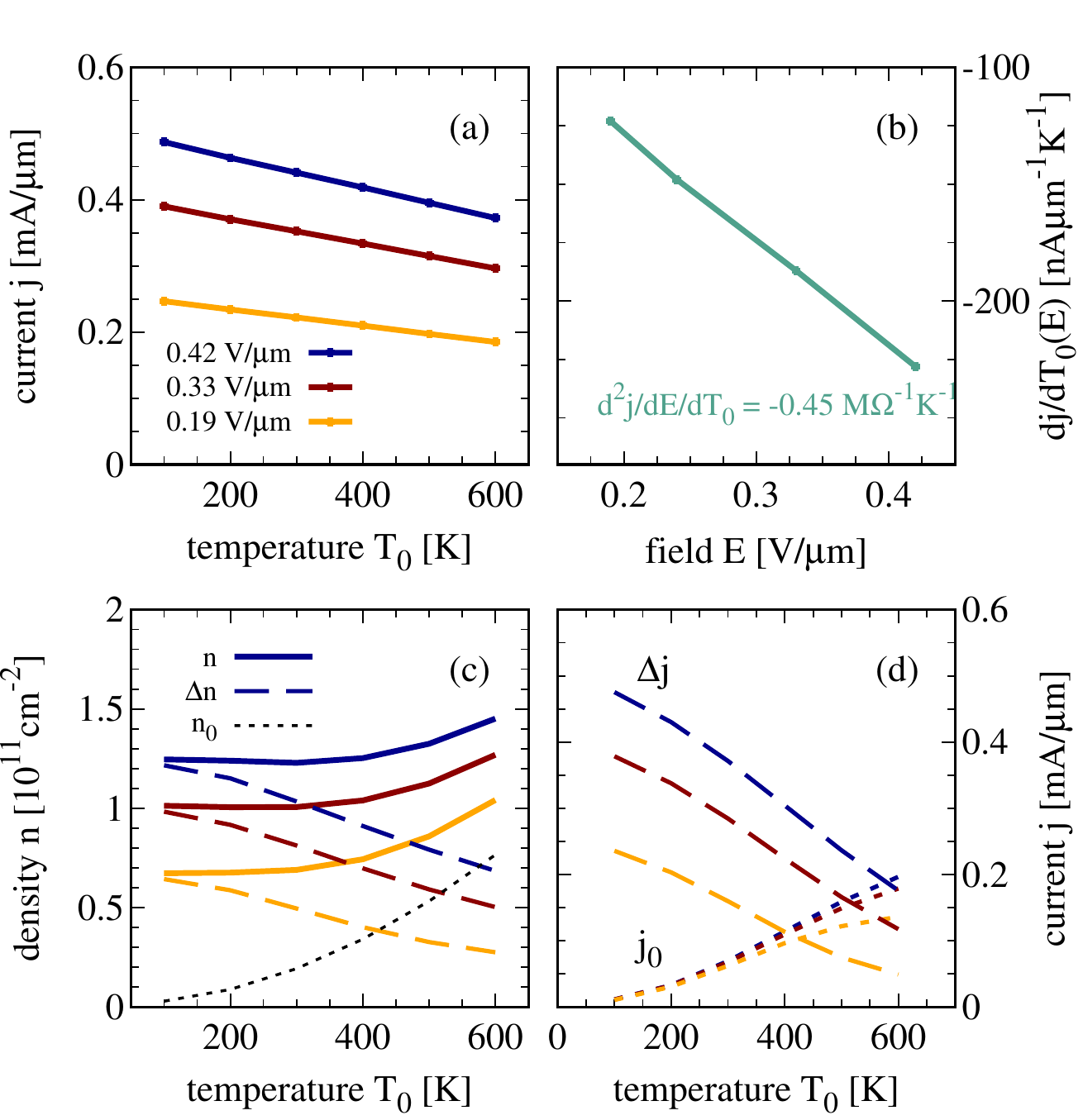} 
\caption{(a) Temperature dependence of the dark current for different electrical fields. (b) The field dependence of $dj/dT_{0}$ is determined by the dark carrier multiplication (dCM) resulting in an increase of the carrier density $n$. Temperature-dependent components of (c) the carrier density $n = n_{0} + \Delta n$ and (d) the current $j = j_{0} + \Delta j$.}
\label{fig:field} 
\end{figure}
In the absence of optical excitation, the dark current $j_{dark}$ decreases linearly with $T_{0}$ for different electric fields, cf. Fig. \ref{fig:field}(a). Due to the linear scaling, $dj_{dark}/dT_{0}$ is constant for a given $E$, and its electric field dependence is illustrated in Fig. \ref{fig:field}(b). To understand this behaviour, we consider the two key factors determining the current density $j_{dark}$: (i) the number of charge carriers which is reflected by the carrier density $n$, and (ii) many-particle scattering which redistributes the charge carriers and thereby counteracts their acceleration in the electric field. The second factor is the average velocity of charge carriers in the electric field. It is denoted as drift velocity $v_{d}$ and results from an equilibrium of the many-particle scattering counteracting the field-induced acceleration of charge carriers. In graphene, the two factors are not independent of each other, since efficient Auger scattering has an impact on the carrier density. As was thoroughly investigated in Ref. \onlinecite{Jago_darkCM_2017}, Auger scattering results in a carrier density increase when graphene is exposed to an external in-plane electric field. We state that this dark carrier multiplication (dCM) is the prevailing mechanism behind the observed behaviour of the numerically calculated $j_{dark}$.

Therefore, it is instructive to separate the carrier density into the initially available carrier density due to thermally distributed charge carriers $n_{0}$ and the carrier density change induced by Auger scattering $\Delta n$. The temperature dependence of $n_{0}$ can be obtained from the Fermi distribution by integrating over $\mathbf{k}$, which yields $n_{0} =\frac{\pi k^{2}_{B}}{3\,\hbar^{2} v_{F}}T_{0}^2$, cf. the dashed black line in Fig. \ref{fig:field}(c). Adding the carrier density increase induced by dCM for the respective electric field (dashed coloured lines in Fig. \ref{fig:field}(c)), the total carrier density is obtained (solid lines). Based on $n_{0}$, $\Delta n$, and $v_{d}$, we calculate $j_{0}$ and $\Delta j$ using the relation $j=e_{0}v_{d}n=e_{0}v_{d}(n_{0}+\Delta n) = j_{0}+\Delta j$. The result is shown in Fig. \ref{fig:field}(d) and demonstrates that $j_{0}(T_{0},E)$ and $\Delta j(T_{0},E)$ reflect the behaviour of $n_{0}(T_{0},E)$ and $\Delta n(T_{0},E)$, respectively. However, there is a small distortion due to $v_{d}(T_{0},E)$ resulting in a slower increase of $j_{0}$ for larger temperatures.
The latter increases with temperature showing the same scaling as $n_{0}$ (i.e. $j_{0}\propto T^2_{0}$) for low temperatures and large electric fields but exhibiting a saturation as the temperature increases or the electric field becomes weaker. At the same time, $\Delta j$ shows the same behaviour as $\Delta n$ and $j$: it becomes smaller with temperature but becomes larger with the electric field, and the slope $dj/dT_{0}$ becomes steeper for higher fields. As a result, we can conclude that the temperature and the electric field dependence of the total current density $j$ is governed by dCM, at least for not too high temperatures and not too low electric fields. Since the dark CM is strong, more carriers are created in the conduction band which increases the resistivity such that the conductivity/current decays slower with temperature ($\propto -T$), therefore the temperature dependence deviates from the Gr\"uneisen formula ($\propto 1/T$).

%====================================================================================================================================================
\bigskip
%====================================================================================================================================================
\begin{figure}[!t]
\centering
\includegraphics[width=0.9\columnwidth]{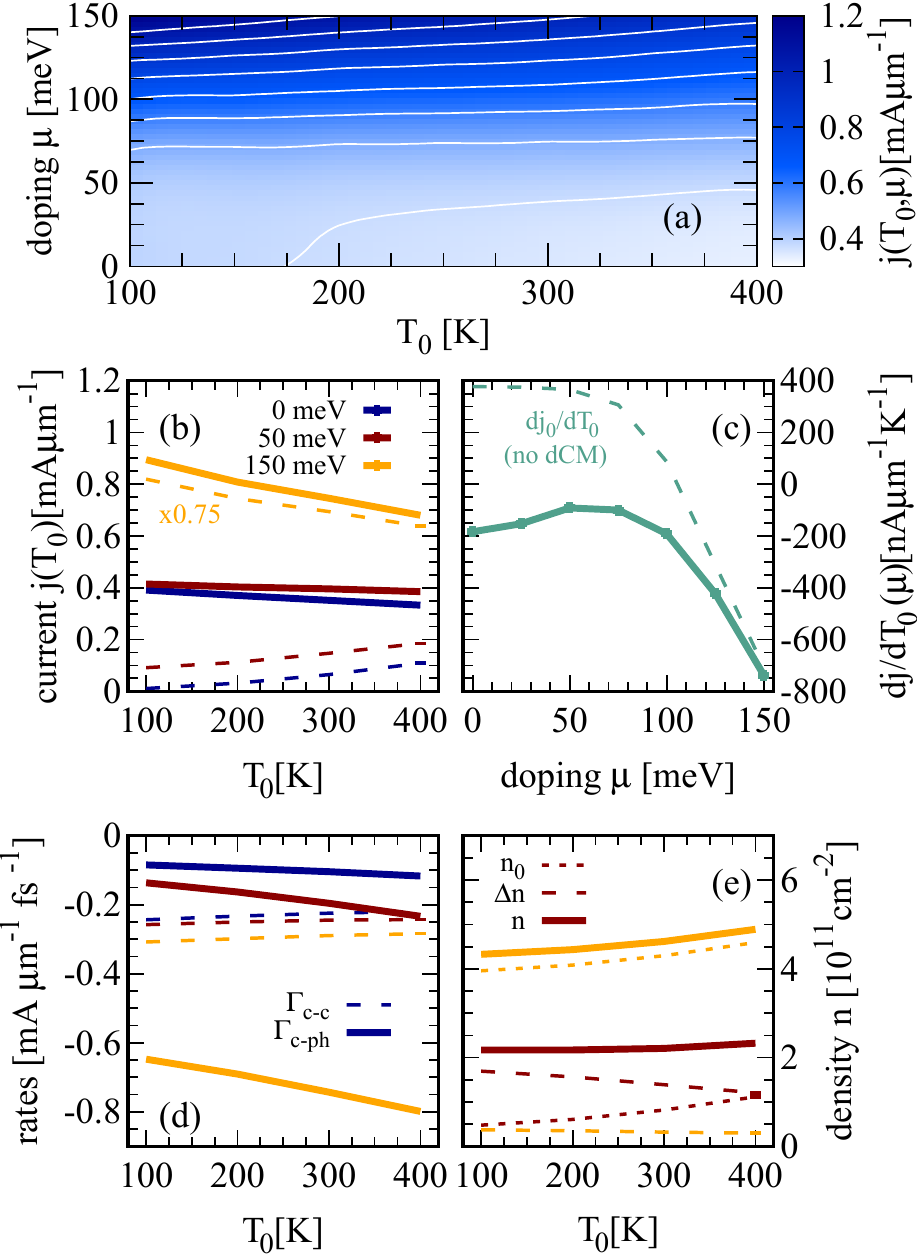} 
\caption{(a) Current for different doping values and temperatures for the fixed electric field of $0.33\,\text{V}/\mu\text{m}$. (b) Temperature dependence of current for different doping values. The dashed lines represent the current without dCM. (c) Doping dependence of $dj_{0}/dT_{0}$ (without dCM) and $dj/dT_{0}$ (with dCM). (d) Scattering rates for different doping values (according to (b)) representing the current changes due to carrier-carrier $\Gamma_\text{c-c}$ and carrier-phonon scattering $\Gamma_\text{c-ph}$. It demonstrates the increasing impact of carrier-phonon scattering at higher doping. (e) Carrier density $n = n_{0} + \Delta n$ for a doping value of $50\,\text{meV}$ (red) and $150\,\text{meV}$ (orange).
}
\label{fig:doping} 
\end{figure}
Since the bolometric effect is known to dominate the photoresponse of biased graphene \cite{Freitag2013a,Mittendorff2013_thzdetector,Mittendorff2015_Photodetector,Schuler2016}, it is crucial to investigate $dj_{dark}/dT_{0}$ for a finite doping. Note that our equations of motion are electron-hole symmetric and thus the same current density is found when the sign of the doping is reversed. An investigation of $j_{dark}$ for different temperatures and doping reveals that it can be significantly increased in doped samples, cf. Fig. \ref{fig:doping}(a). This is a direct consequence of the increased initial carrier density in the presence of doping. Furthermore, the scaling of $j_{dark}$ with the temperature remains linear for finite doping, cf. Fig. \ref{fig:doping}(b), with the slope $dj_{dark}/dT_{0}(\mu)$ showing an interesting behaviour: it increases up to $\mu \approx 60\,\text{meV}$ before it strongly decreases, cf. Fig. \ref{fig:doping}(c).

To understand this behaviour we define current rates to separate the change of current induced by carrier-carrier and carrier-phonon scattering via
\begin{align}
 \frac{d\mathbf{j}}{dt}\bigg|_{\text{c-c/c-ph}} = -\frac{g\, e_{0}v_{\text{F}}}{A}\sum_{\bf{k}\lambda}\mathbf{e}_{\bf{k}}\,\dot{\rho}^{\lambda}_{\bf k}\bigg|_{\text{c-c/c-ph}}
\end{align}
with $\dot{\rho}^{\lambda}_{\bf k}\big|_{\text{c-c/c-ph}} = \Gamma_{\mathbf{k}\lambda,\text{c-c/c-ph}}^{\text{in}}\,\big(1-\rho_{\mathbf{k}}^{\lambda}\big)-\Gamma_{\mathbf{k}\lambda,\text{c-c/c-ph}}^{\text{out}}\,\rho_{\mathbf{k}}^{\lambda}$.

We first consider the case without the dCM, cf. the dashed lines in Figs. \ref{fig:doping}(b) and \ref{fig:doping}(c). At zero doping, $j_{0}$ increases with the temperature (dashed lines in Fig. \ref{fig:field}(c)), hence the value of $dj_{0}/dT_{0}(\mu=0)$ is positive. For finite doping, $j_{0}$ increases due to an enhanced carrier density (dotted black line in Fig. \ref{fig:field}(d) and dotted lines in Fig. \ref{fig:doping}(e)). At the same time, $dj_{0}/dT_{0}$ decreases monotonically with the amount of doping. This is caused by an amplification of the carrier-phonon scattering rate, which has a negative influence on the current density growing with the temperature (solid lines in Fig. \ref{fig:doping}(d)). At the same time, the impact of carrier-carrier scattering stays nearly constant (dashed lines in Fig. \ref{fig:doping}(d)). Moreover, the enhanced carrier-phonon scattering is more sensitive to temperature changes.

Starting from $j_{0}$, the behaviour of the  complete current density $j$ (solid lines in Fig. \ref{fig:doping}(b) and \ref{fig:doping}(c)) is readily understood by adding the contribution of the dCM to $j_{0}$. The impact of the dCM is illustrated via $\Delta n$ which is just proportional to $\Delta j$. The carrier density increase $\Delta n$ is reduced with rising temperatures (dashed lines in Fig. \ref{fig:field}(d)), and it vanishes at large doping values (Fig. 3(e)). The collinear Auger-type scattering does not longer occur near the Dirac point, where it bridges the valence and the conduction band, but  it becomes a regular intraband scattering channel at higher energies (in presence of doping). This explains the maximum of $dj/dT_{0}$ at intermediate doping values (around $50\,\text{meV}$): For vanishing doping,  dCM is maximal and results in a strong decrease of $dj/dT_{0}$. As the doping increases, the impact of the dCM becomes smaller and this decrease has a higher impact than the decrease of $dj_{0}/dT_{0}$ (caused by the enhancement of carrier-phonon scattering) resulting in an increase of $dj/dT_{0}$. At even higher doping values, the impact of carrier-phonon scattering rises drastically and consequently $dj/dT_{0}$ eventually decreases. The reason for the enhanced carrier-phonon scattering is that the carrier distribution for higher doping values is much broader and enables relaxation via optical phonons such that the current becomes more sensitive to temperature changes and therefore also the conductivity shows a sizeable variation with temperature.

%====================================================================================================================================================
\subsection*{Optically induced change of temperature}
%====================================================================================================================================================
\begin{figure}[!t]
\begin{centering}
\includegraphics[width=0.95\columnwidth]{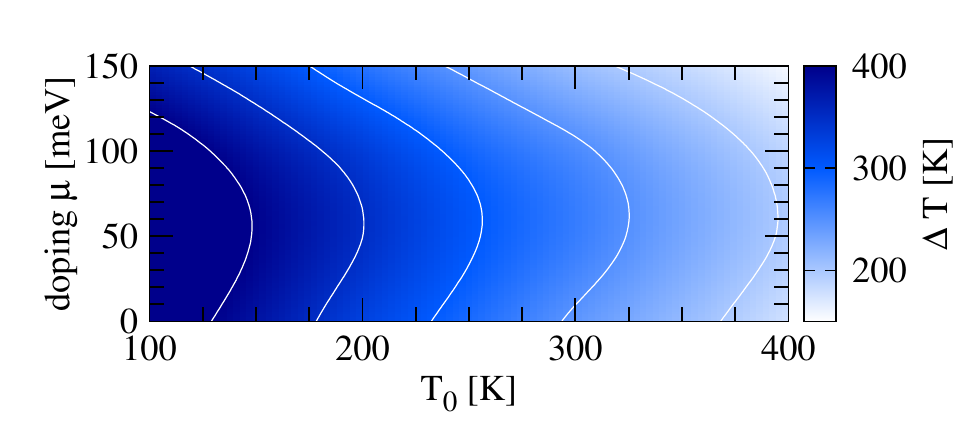} 
\par\end{centering}
\caption{Maximal temperature increase $\Delta T$ (with respect to the initial temperature $T_{0}$) after optical excitation as a function of doping $\mu$ and $T_{0}$ at a fixed pump fluence of $1\mu \text{J/cm}^{2}$.}
\label{fig:optical-excitation}
\end{figure}
Now, we consider the situation without electric field but with optical excitation. The graphene layer is excited with a laser pulse of the energy $\hbar\omega_{L}=0.6\,\text{eV}$ and with a pump fluence of $1\,\mu\text{Jcm}^{-2}$, this low fluence reduces the impact of nonlinear optical response. The excited carriers relax towards lower energies and redistribute the carrier distribution via carrier-carrier and carrier-phonon scattering such that a quasi-equilibrium distribution is reached. The latter corresponds to a hot Fermi distribution with an usually much larger temperature than in equilibrium situation.

The dependence of the temperature change $\Delta T$ on initial temperature and doping is depicted in Fig. \ref{fig:optical-excitation}. By increasing the doping (up to $\approx 70\,\text{meV}$) the temperature of the electronic system increases, while for even larger doping values, the temperature decreases again.

In the valence band the phase space (for holes) between optical excitation and Fermi level becomes larger for increasing doping. The carrier scattering (mainly collinear impact excitation-type scattering) forms a smooth transition from the excitation energy to the Fermi level resulting in a broader carrier distribution and therefore a higher temperature. 

The decrease of temperature for doping values above $70\,\text{meV}$ can be ascribed to the significant increase of the charge carrier density. As a result, the energy induced by optical excitation, can be distributed to more carriers and therefore the energy per carrier decreases, which corresponds to a reduced temperature.

In the case without doping we estimate the temperature increase due to optical excitation with carrier density $n$ and total carrier energy $\varepsilon$. Assuming Fermi distributions we get $n = k^{2}_{B} \pi/(3\hbar^{2}v^{2}_{F})\,T^{2}_{0} = c_{1} T^{2}_{0}$ and $\varepsilon = 6 k^{3}_{B}\zeta(3) /(\hbar^{2}v^{2}_{F}\pi)\,T^{3}_{0} = c_{2} T^{3}_{0}$ and therefore we can express the density in terms of the energy $n = c_{1}(\varepsilon/c_{2})^{2/3}$. Adding the energy of the optical pulse $\varepsilon_{opt}$ results in $n = c_{1}(T^{3}_{0} + \varepsilon_{opt}/c_{2})^{2/3}$ with final temperature $T = (T^{3}_{0} + \varepsilon_{opt}/c_{2})^{1/3}$ and the temperature increase is given by $\Delta T = T-T_{0}$. When we evaluate this analytic formula we obtain the same behaviour as our numerical calculations but with a slightly higher temperature ($\approx 50\,$K). This discrepancy is a result of relaxation processes already during the excitation.

%====================================================================================================================================================
 \subsection*{Bolometric photocurrent}
%====================================================================================================================================================
Combining the two contributions $dj/dT_{0}$ (Fig. \ref{fig:doping}(c)) and $\Delta T$ (Fig. \ref{fig:optical-excitation}) determining the bolometric photocurrent $j_{\text{bolo}}$, we reveal its doping, temperature, field and pump fluence dependence. 

Regarding doping and temperature (Fig. \ref{fig:sketch}(c)), the bolometric photocurent shows essentially the same behaviour as $dj/dT_{0}$ with an additional temperature dependent parabolic distortion due to $\Delta T (T_{0})$. The bolometric photocurrent is negative with a minimal amplitude at a doping of $50\,\text{meV}$.  The current is slightly larger at lower doping values and becomes significantly increased for doping values larger then 100, meV. This reflects  exactly the behaviour of $dj/dT_{0}$ shown in  Fig. \ref{fig:doping}(c). 
The bolometric photocurrent is enhanced at lower temperatures (Fig. \ref{fig:sketch}(c)), which results from a larger photo-induced temperature change $\Delta T$, as shown in Fig. \ref{fig:optical-excitation}.

\begin{figure}[!t]
 \begin{centering}
 \includegraphics[width=0.95\columnwidth]{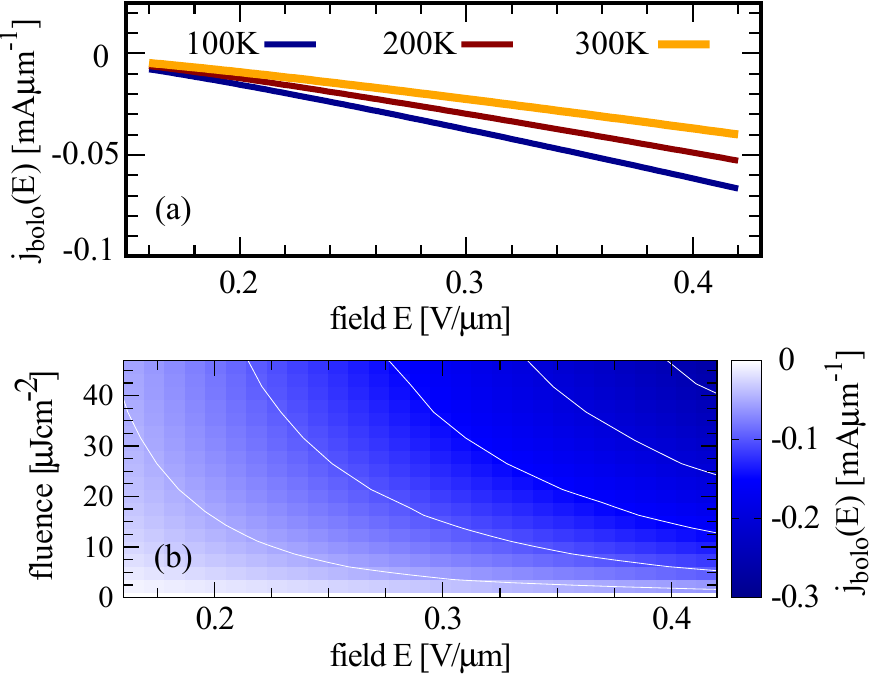} 
 \par\end{centering}
 \caption{The bolometric photocurrent in doped graphene ($50\,\text{meV}$) evaluated for (a) different electrical fields and initial temperatures $T_{0}$ at a fixed pump fluence of $1\mu \text{J/cm}^{2}$ and (b) different pump fluences and fields at a fixed initial temperature $T_{0}=300\text{K}$.}
\label{fig:optic-fluence} 
\end{figure}

Furthermore, $j_{\text{bolo}}$ scales linearly with the electric field (Fig. \ref{fig:optic-fluence}(a)), which is a consequence of the increasing dark carrier multiplication (Fig. \ref{fig:field}(b)). Since the dCM is strongly reduced in doped graphene, the bolometric photocurrent is almost independent of the initial temperature for small electric fields (Fig. \ref{fig:optic-fluence}(a)). For sufficiently strong fields, the field-induced distortion of the initial carrier distribution becomes so large that dCM becomes possible again. 

Finally, the  bolometric photocurrent becomes considerably enhanced, when the pump fluence of the excitation pulse is increased (Fig. \ref{fig:optic-fluence}(b)).  Changing the pump fluence from $1\,\mu\text{J/cm}^{2}$ to $50\,\mu\text{J/cm}^{2}$ at room temperature, $j_{\text{bolo}}$ can be enhanced by almost a factor of $7$. This is due to the fluence-induced increase of the thermalized electron temperature. The larger the optical excitation, the more energy is transferred to the electronic system, and the higher is the temperature of the thermalized quasi-equilibrium distribution. 

%====================================================================================================================================================
\subsection*{Conclusions}
%====================================================================================================================================================
In summary, we provide a microscopic view on the generation of photocurrents in intrinsic and doped graphene in presence of an in-plane electric field. Treating the time- and momentum-resolved interplay of field-induced acceleration of optically excited carriers and carrier-carrier and carrier-phonon scattering on the same microscopic footing, we reveal the microscopic foundation of the bolometric effect in graphene. In particular, we provide an explanation for the experimental observation that the bolometric effect governs the photoresponse of highly doped graphene.  The current in unbiased graphene is determined by the amplification of the carrier density due to Auger scattering. This dark carrier multiplication results in a decrease of the current with temperature corresponding to a negative bolometric photocurrent. The predominant role of the bolometric effect in highly doped graphene can be ascribed to the carrier-phonon scattering. Its increasing temperature sensitivity at high doping results in a strong amplification of the bolometric photocurrent.  

%====================================================================================================================================================
\subsection*{Acknowledgements}
%====================================================================================================================================================
This project has received funding from the European Union's Horizon 2020 research and innovation programme under grant agreement No 696656 (Graphene Flagship). Furthermore, we acknowledge support from the Swedish Research Council (VR). The computations were performed on resources at Chalmers Centre for Computational Science and Engineering (C3SE) provided by the Swedish National Infrastructure for Computing (SNIC).

\end{document}